\documentclass{article}
\usepackage[utf8]{inputenc}
\usepackage{url} % include urls
\usepackage{hyperref} % use hyperlinks
\usepackage[affil-it]{authblk}
\usepackage[colorinlistoftodos]{todonotes} % notes in the pdf
\usepackage{color, colortbl}
\usepackage{multirow}
\usepackage{graphicx}
\usepackage{siunitx}
\usepackage{booktabs}

\usepackage[
   backend=biber,
   style=ieee, %numeric-comp, % use 'numeric' in order not not combine citations
   sorting=none,
   natbib=true,
   maxcitenames=1,
   mincitenames=1,
   url=true, 
   doi=false,
   eprint=false
]{biblatex}
\newcommand{\citest}[1]{\citefield{#1}{shorttitle}}
\newcommand{\citetcia}[2]{\citefield{#1}{shorttitle} \cite{#1,#2}}

\providecommand{\keywords}[1]
{
  \textbf{\textit{\\Keywords: }} #1
}

\definecolor{VeryLightGray}{gray}{0.95}
\definecolor{LightGray}{gray}{0.85}
\definecolor{Gray}{gray}{0.75}

\title{Deep Learning Based HPV Status Prediction for Oropharyngeal Cancer Patients}

\author[1, 2, 3]{D.M. Lang}
\author[1, 2, 4]{J.C. Peeken}
\author[1, 2, 4]{S.E. Combs}
\author[2, 3]{J.J. Wilkens}
\author[1, 2]{S. Bartzsch}

\affil[1]{Institute of Radiation Medicine, Helmholtz Zentrum M\"unchen, Munich, Germany}
\affil[2]{Department of Radiation Oncology, School of Medicine and Klinikum rechts der Isar,
         Technical University of Munich (TUM), Munich, Germany}
\affil[3]{Physics Department, Technical University of Munich, Garching, Germany}
\affil[4]{Deutsches Konsortium f\"ur Translationale Krebsforschung (DKTK), Partner Site Munich}
\affil[ ]{\fontsize{8}{12}\fontfamily{lmtt}\selectfont{\{daniel.lang, jan.peeken, stephanie.combs, wilkens, stefan.bartzsch\}@tum.de}}

\date{}

\begin{document}

\maketitle

% -------------------------------- Abstract  --------------------------------------%

% should be 200 words max
\begin{abstract}
Infection with the human papillomavirus (HPV) has been identified as major risk factor for oropharyngeal cancer (OPC).
HPV-related OPCs have been shown to be 
more radiosensitive and to have a reduced risk for cancer related death.
Hence, histological determination of HPV status of cancer patients depicts an essential diagnostic factor.

We investigated the ability of deep learning models for imaging based HPV status detection.
To overcome the problem of small medical datasets we used a transfer learning approach.
A 3D convolutional network pre-trained on sports video clips was fine tuned such that full
3D information in the CT images could be exploited.

The video pre-trained model was able to differentiate HPV-positive from HPV-negative cases
with an area under the receiver operating characteristic curve (AUC) of $0.81$ for an external test set.
In comparison to a 3D convolutional neural network (CNN) trained from scratch and a
2D architecture pre-trained on ImageNet the video pre-trained model performed best.

Deep learning models are capable of CT image based HPV status determination.
Video based pre-training has the ability to improve training for 3D medical data
but further studies for verification are needed.

\\
~
\end{abstract}

% -------------------------------- Intoduction --------------------------------------%

\section{Introduction}

Tumor response to radiation therapy depends on multiple factors, such as histology, tumor stage, or molecular aberrations.
The increasing availability of diagnostic biological and imaging data in radiation oncology
opens new perspectives to personalize treatment and thereby improve treatment outcome.
Machine learning techniques in combination with high-dimensional personalized ``-omics'' datasets have been shown
to be powerful tools for prognostic and predictive assessment of therapeutic efficacy {\citep{Coates2016, Naqa2017, Kumar2012}}.
Radiomics refers to the extraction of information from radiological images by applying hand-crafted filters
on preselected regions of interest \citep{Peeken2020a}.
\citet{Segal2007} demonstrated that features in radiological images can be used to reconstruct
the majority of the tumor genetic profile.
Radiomics data successfully predicted overall survival \citep{Fave2017,Spraker2019},
metastases development \citep{Huynh2016,Peeken2019a} or histological properties \citep{Peeken2019,Peeken2020}
and may be used as decision support system in clinical practice. 

Although radiomics has proven its potential in medical image analysis,
deep learning was shown to be clearly superior in most other computer vision tasks.
Deep learning features an end-to-end training without the need to design hand-crafted filters.
However, training requires considerably larger patient numbers than typically available in clinical applications.
Large 3D-data poses another challenge on the application of deep learning in medical image analysis.

In this work, we investigate deep learning on diagnostic CT-images as a tool to attribute oropharyngeal cancer (OPC)
to a human papillomavirus virus (HPV) driven oncogenesis.
HPV has been identified oncogenic for several cancer sites \citep{IARC2007}. Chronic HPV infections
may also lead to OPC development \citep{Mork2001}.
While smoking and alcohol consumption, two well established risk factors for OPC,
have notably declined in North America and Northern Europe \citep{Ng2014},
infections with HPV have increased and lead to growing incidence rates of OPC \citep{IARC2007}.
\citet{Plummer2016} estimated in 2016 that around \SI{31}{\%} of OPC cases globally are caused by HPV.
However, OPC patients with a positive HPV status show a \SI{74}{\%} reduced risk of cancer related death \citep{Gillison2000}
and HPV-positive tumors are more radiosensitive than HPV-negative tumors.
Hence, determination of the HPV status has become an essential diagnostic factor and dose de-escalation studies,
such as the ECOG 1308 trial \citep{Marur2017}, seek to reduce therapy induced side effects for HPV-positive OPC patients. 

Different histological methods exist to determine the HPV status.
Most frequently used is the detection of HPV induced overexpression of p16$^\text{INK4a}$ by immunohistochemistry
with a reported sensitivity of \SI{>90}{\%} and a specificity of \SI{>80}{\%} \cite{schache2011evaluation, jordan2012validation}.
Non-invasive determination of the HPV status based on routinely acquired radiological imaging may simplify the diagnostic process and reduce costs. 

Radiomics approaches to determine the HPV status achieved areas under the receiver operating characteristic curve (AUC) of about \SIrange{70}{80}{\%},
when tested on external data sets \citep{bogowicz2020privacy, huang2019development}.
\citet{fujima2020prediction} trained a 2D convolutional neuronal network (CNN) on FDG-PET images to classify  HPV status in OPC patients
and achieved an AUC of \SI{83}{\%} using a transfer learning approach based on natural images from the ImageNet database \citep{ILSVRC15}.
But, they did not test their data on an external cohort and excluded images containing severe motion artifacts
and tumors with diameter sizes below \SI{1.5}{cm}.
Our study was based on 4 different publicly available data sets of the TCIA archive \cite{clark2013cancer}.
Transfer learning facilitated deep learning on the relatively small data set size on a CNN
derived from the C3D classification network \citep{Tran_2015_ICCV},
with weights pre-training on the Sports-1M data set \citep{ILSVRC15}.
\citet{Hussein_2017} used C3D to initialize a multi-task learning approach for lung nodule risk stratification.
However, their network depended on additional information such as tumor sphericity and texture while we
trained our network in a simple end-to-end fashion.

% --------------------------------   Methods   --------------------------------------%

\section{Material and Methods}

% --------------------------- Data --------------------------------- %

\subsection{Data}

% TCIA data

Head and neck cancer collections
\citetcia{tciaOPC}{kwan2018radiomic}, \citetcia{tciaHNSCC}{grossberg2018imaging},
\citetcia{tciaHNPC}{vallieres2017radiomics} and \citetcia{tciaHN1}{aerts2014decoding} 
of the publicly accessible TCIA archive \cite{clark2013cancer} were mined for appropriate cases.
Inclusion criteria were: oropharyngeal subtype, existence of a pre-treatment CT image with respective segmentation
of the gross tumor volume (GTV) and detected HPV status.
Only the center point of the GTV was used to cut the images to smaller size,
i.e. no exact delineation of the tumor was needed.
In total this led to 850 individual oropharyngeal cancer patients (Table \ref{tab_meta}).

\newcolumntype{G}{>{\columncolor{Gray}}c}
\newcolumntype{L}{>{\columncolor{LightGray}}c}
\newcolumntype{V}{>{\columncolor{VeryLightGray}}c}

\begin{table}[h!]
   \centering
   \scalebox{0.6}{
   \begin{tabular}{| l r | l l l l |}
      \multicolumn{2}{l}{}             & \multicolumn{2}{L}{training set} & \multicolumn{1}{V}{validation set} & \multicolumn{1}{L}{test set} \\
      \multicolumn{2}{l}{}              & \textbf{OPC}       & \textbf{HNSCC}     & \textbf{HN PET-CT} & \multicolumn{1}{l}{\textbf{HN1}}          \\
      \hline
      \multicolumn{2}{|l|}{Patients}      & 412                & 263                & 90                    & 80                    \\  
      \multicolumn{2}{|l|}{HPV: pos/neg}  & 290/122            & 223/40             & 71/19                 & 23/57                 \\  
      \hline
      \multicolumn{2}{|r|}{
         \hspace{3cm}HPV status}          &                    &                    &                       &                       \\  
      \multicolumn{2}{|l|}{Age}           &                    &                    &                       &                       \\  
      \multicolumn{2}{|r|}{pos}           & 58.81 (52.00-64.75)& 57.87 (52.00-64.00)& 62.32 (58.00-66.00)   & 57.52 (52.00-62.50)   \\  
      \multicolumn{2}{|r|}{neg}           & 64.82 (58.00-72.75)& 60.02 (54.50-67.25)& 59.11 (49.50-69.50)   & 60.91 (56.00-66.00)   \\  

      \multicolumn{2}{|l|}{
         Sex: Female/Male}                &                    &                    &                       &                       \\  
      \multicolumn{2}{|r|}{pos}           & 47/243             & 32/191             & 14/56                 & 5/18                  \\  
      \multicolumn{2}{|r|}{neg}           & 34/88              & 15/25              & 4/15                  & 12/45                 \\  

      \multicolumn{2}{|l|}{
         T-stage: T1/T2/T3/T4}            &                    &                    &                       &                       \\  
      \multicolumn{2}{|r|}{pos}           & 46/93/94/57        & 60/93/41/29        & 10/37/15/9            & 4/8/9/8               \\  
      \multicolumn{2}{|r|}{neg}           & 9/35/43/35         & 6/12/12/10         & 3/4/8/4               & 9/16/9/23             \\  

      \multicolumn{2}{|l|}{
         N-stage: N0/N1/N2/N3}            &                    &                    &                       &                       \\  
      \multicolumn{2}{|r|}{pos}           & 33/22/215/20       & 19/30/170/4        & 11/10/47/3            & 6/2/15/0              \\  
      \multicolumn{2}{|r|}{neg}           & 36/16/62/8         & 5/2/31/2           & 2/1/13/3              & 14/10/31/2            \\  

      \multicolumn{2}{|l|}{
         Tumor size $[cm^{3}]$}           &                    &                    &                       &                       \\  
      \multicolumn{2}{|r|}{pos}           & 29.35 (10.52-37.78)& 11.78 (3.94-14.04) & 34.63 (14.91-41.77)   & 23.00 (10.83-34.29)   \\  
      \multicolumn{2}{|r|}{neg}           & 36.99 (15.72-45.35)& 23.57 (5.80-22.85) & 35.09 (17.32-47.82)   & 40.19 (11.77-54.42)   \\  
      \hline
      \multicolumn{2}{|l|}{
      transversal voxel spacing $[mm]$}   & 0.97 (0.98-0.98)   & 0.59 (0.49-0.51)   & 1.06 (0.98-1.17)      & 0.98 (0.98-0.98)      \\  
      \multicolumn{2}{|l|}{
      longitudinal voxel spacing $[mm]$}  & 2.00 (2.00-2.00)   & 1.53 (1.00-2.50)   & 2.89 (3.00-3.27)      & 2.99 (3.00-3.00)      \\  
      manufacturer \hspace{0.5cm} &       &                    &                    &                       &                       \\              
       & \multicolumn{1}{l|}{GE Med. Sys.}& 272                & 238                & 45                    & 0                     \\
       & \multicolumn{1}{l|}{ Toshiba}    & 138                & 3                  & 0                     & 0                     \\
       & \multicolumn{1}{l|}{ Philips}    & 2                  & 12                 & 45                    & 0                     \\
       & \multicolumn{1}{l|}{CMS Inc.}    & 0                  & 0                  & 0                     & 43                    \\
       & \multicolumn{1}{l|}{Siemens}     & 0                  & 4                  & 0                     & 37                    \\
       & \multicolumn{1}{l|}{other}       & 0                  & 6                  & 0                     & 0                     \\
      \hline
   \end{tabular}
   }   
   \caption{
      Patient information for the different cohorts. Continuous variables are represented by mean values
      and ranges by $(q_{25} - q_{75})$, with $q_{25}$ and $q_{75}$ being the $25^{th}$ and $75^{th}$ percentiles, respectively.
   }   
   \label{tab_meta}
\end{table}

To ensure generalizability to images from institutions and scanners not seen during training,
testing on external data is inevitable in a medical setting \cite{park2018methodologic}.
Hence, for training, validation and testing independent data sets were employed.
The \citest{tciaOPC} and \citest{tciaHNSCC} data sets
were combined and used as a training set, since these two cohorts contained the most cases.
Due to its variety in cases with data coming from 4 different institutions
\citest{tciaHNPC} was employed for validation. 
The validation set is used for selection of the final model weights,
a versatile validation set therefore supports selection of a model applicable to a broad kind of settings.
The \citest{tciaHN1} data set was used as test set.

The \citest{tciaOPC} and \citest{tciaHN1} data sets provided the HPV status
tested by immunohistochemical (IHC) based p16 staining.
A combination of p16 IHC and/or HPV DNA in situ hybridization was used in the \citest{tciaHNSCC} data.
For \citest{tciaHNPC} testing methods were not reported.

We resampled all CT images to an isotropic voxel size of $1mm^3$.
Voxel values, given in Hounsfield units, were cropped at $-250$ HU and $250$ HU
and linearly rescaled to integer values ranging from $0$ to $255$.

% ------------------------- Deep Learning -------------------------- %

\subsection{Deep Neural Network}

Transfer learning is commonly used to overcome the problem of small data set sizes.  
A widely applied approach uses the ImageNet data set \cite{ILSVRC15}, consisting of natural images, for pre-training.
However, CT images are 3 dimensional and therefore pre-training should be performed on 3 dimensional data.
We tested the capability of video data based pre-training defining the time axis as the 3$^{rd}$ dimension.

To avoid long training times and obtain a reproducible starting point
we used the already trained video classification network C3D \cite{Tran_2015_ICCV, c3dweb} as
pre-trained base line model.
C3D processes video input in a simple 3D convolutional manner,
i.e. all three input dimensions are handled in the same way.
Convolutional layers are followed by Max-Pooling layers ending with three
densely connected layers and a softmax activation layer.
C3D was trained to predict labels for video clips of the Sport-1M data set \cite{karpathy2014large} which
contains 1.1 million videos of sport activities belonging to one of 487 categories.
For training, $16$ image frames per video clip with a size of $112 \times 112$ were used,
i.e. input dimension was given by $16 \times 112 \times 112 \times 3$.

\begin{figure}
   \centering
   \includegraphics[width=\textwidth]{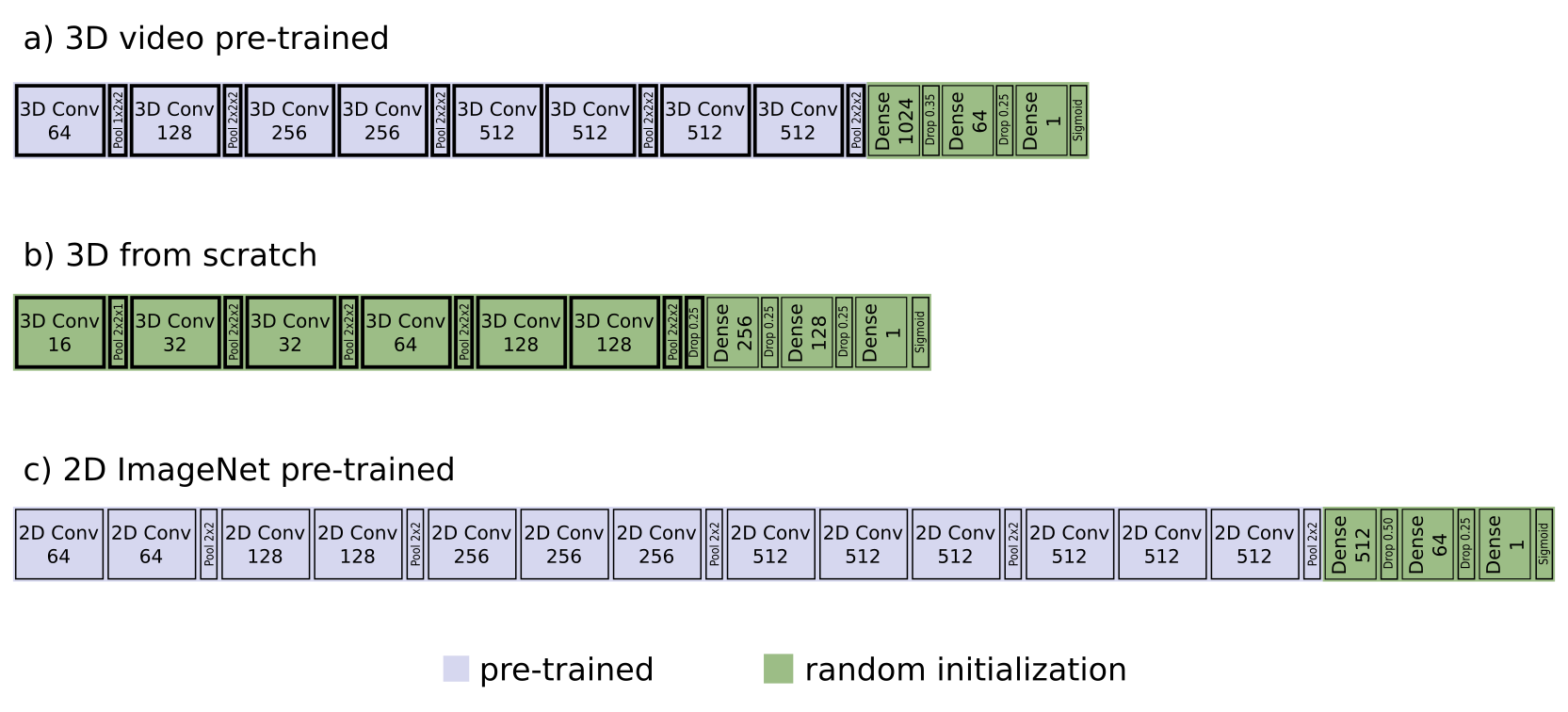}
   \caption{The convolutional neural networks. With a) the network pre-training on the video data of the 1M Sports data set.
            All convolutional layers (Conv) were kept fix while the fully connected layers (Dense) were replace by randomly initialized ones.
            b) shows the CNN trained with all layers randomly initialized.
            Network size was reduced in comparison to a), Max-Pooling (Pool) was applied earlier
            and dropout (Drop) was already used in the last convolutional layer.
            c) the 2D architecture, pre-trained on ImageNet. As base line model VGG16 was used, all convolutional layers were kept fix,
            dense layers were replaced and randomly initialized.
   }
   \label{fig_network}
\end{figure}

All densely connected layers were removed from the network and
weights of all convolutional layers were frozen during training.
New, randomly initialized, densely connected layers were then added after the last convolutional layer,
resulting in a single output neuron followed by a sigmoid activation layer.
All dense layers were followed by a ReLU activation layer and a dropout layer.
The best model consisted of two dense layers with size $1024$ and $64$, with a dropout rate of $0.35$ and $0.25$,
respectively (Figure \ref{fig_network}a).

Weighted binary cross entropy was chosen as loss function with the weights set such that both classes contributed equally.
The Adam optimizer was used with a learning rate of $10^{-4}$ and the batch size was $16$.

As input a single image of size $112 \times 112 \times 48$ was cut from each of the CT scans and then rearranged to fulfill the input requirements,
i.e. 3 consecutive layers were fed to the color channels resulting in information about $16$ of those combined layers in longitudinal direction.
The data augmentation techniques applied included:
flipping on the coronal and the sagittal plane, rotation by a multiple of $90^{\circ}$ and
shifting of the GTV center point by a value between $0$ and $7$ pixels in both directions of the transverse plane.

For comparison, we also trained a 3D convolutional network from scratch, i.e. with all weights randomly initialized,
and a 2D network pre-trained on ImageNet.

The general architecture of the model trained from scratch followed that of C3D net.
Only the size of the network was reduced to avoid overfitting.
To do so, Max-Pooling was applied after every convolutional layer, except for the second but last one and
also dropout was already applied in the last convolutional layer.
The final model consisted of convolutional layers with feature map sizes of $16, 32, 32, 64, 128, 128$
followed by densely connected layers of size $256$ and $128$ (Figure \ref{fig_network}b).
Dropout rate for the last convolutional layer and the two dense layers was $0.25$.
All convolutional kernels were chosen to be of size $3 \times 3 \times 3$.
In order to not merge the signal in the time dimension too early
the C3D model used a kernel and stride of size  $2 \times 2 \times 1$ in the very first Max-Pooling layer.
We followed this approach to account for the smaller input size in longitudinal dimension.
All other Max-Pooling kernels were of size $2 \times 2 \times 2$ with a stride of the same shape.
Input images were cut from the CT scans in exactly the same way as before except for the rearrangement of layers,
i.e. input size was $112 \times 112 \times 48$.
Data augmentation was applied as before and also the optimizer and learning rate stayed the same.

For the 2D model the VGG16 architecture of \cite{simonyan2014deep}, pre-trained on ImageNet, was chosen.
One image was cut from each CT scan in the exact same way as for the 3D model trained from scratch.
For training, the images were split into 16 slices of size $112 \times 112 \times 3$ to fit the networks input dimensions,
i.e. three consecutive slices were fed to the color channels of the network.
During testing, the overall prediction score was constructed by calculation of the mean value for all of those 16 slices. 
All dense layers were removed from the model and replaced by randomly initialized layers.
Dense layers were again followed by ReLU activations and dropout layers to finally end in one single
output neuron with a sigmoid activation function.
Weights of all convolutional layers were again kept fix.
The best performing model had a size of $512$ followed by $64$, with a dropout rate of $0.5$ and $0.25$, respectively
(Figure \ref{fig_network}c).
Data augmentation was performed as before.

% --------------------------------  Results  ----------------------------------------%

\section{Results}

All models were trained for $200$ epochs.
Weights of the epoch with the best performing loss were chosen for the respective final model.
Due to the relative small size of our test and validation sets, we chose to train each model with the same
hyper-parameter settings 10 times and report mean results.

The 3D video pre-trained model achieved the highest validation AUC with a mean (min, max) value of 0.73 (0.69, 0.77)
and a corresponding training AUC of 0.95 (0.90, 0.98).
The 3D network trained from scratch reached a slightly less validation AUC of 0.71 (0.67, 0.74),
training AUC was given by 0.83 (0.66, 0.92).
Results for the 2D network pre-trained on ImageNet were given by 0.62 (0.58, 0.64) for validation and
0.78 (0.76, 0.79) for training.

\begin{figure}
   \centering
   \includegraphics[width=\textwidth]{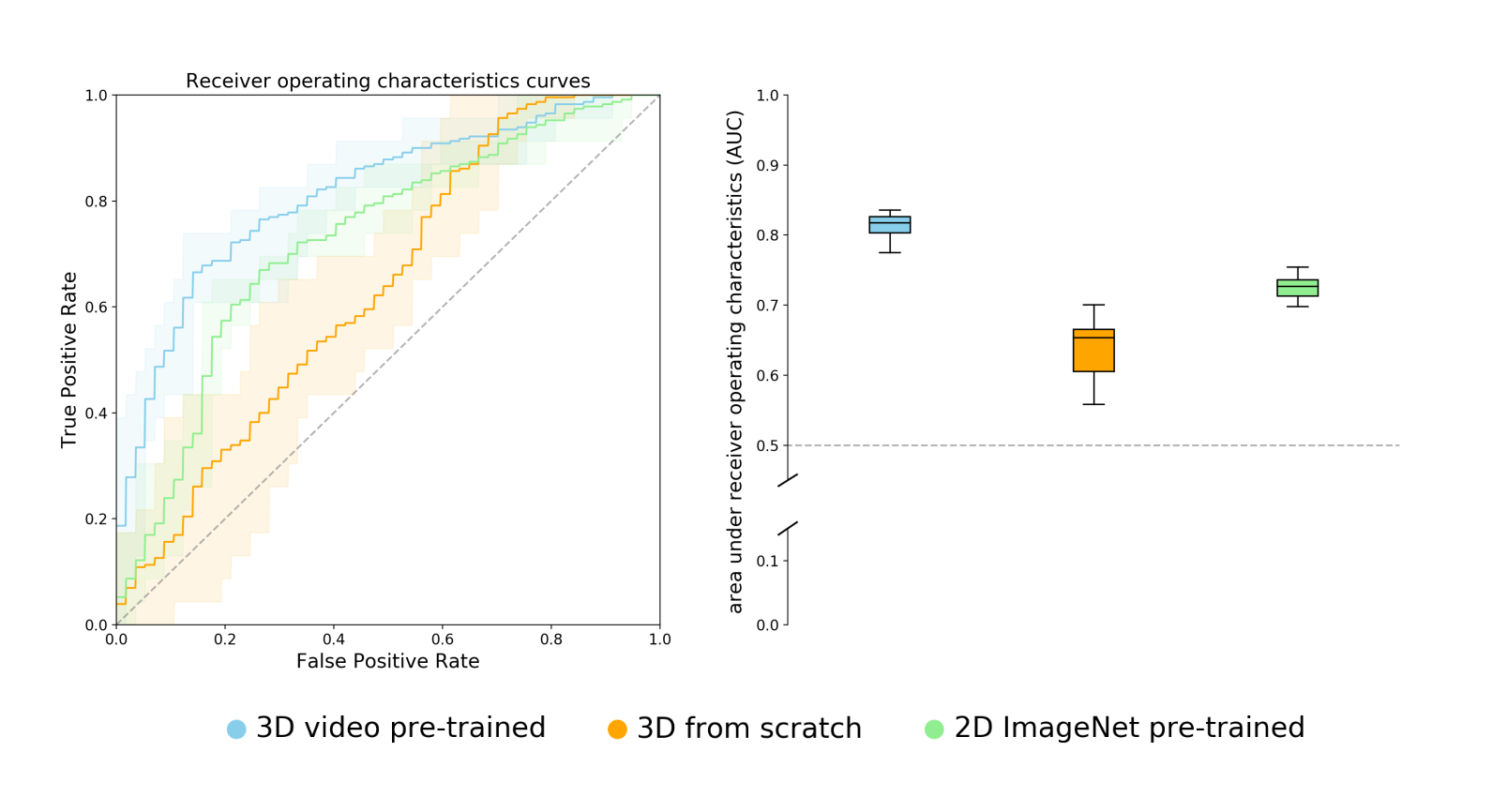}
   \caption{
      Combined receiver operating characteristics (ROC) plots and AUC score box plots
      of the test set for the three different models.
      The ROC plots represent the mean curve of the ten times that models were trained, 
      the shadowed area are represents the minimal and maximal curves.
      Boxes represent lower and upper quartiles, whiskers minimal and maximal values.
      Median (min, max) AUC score for the video pre-trained model was given by  0.82 (0.77, 0.84).
      The 3D network trained from scratch and the 3D ImageNet pre-trained model reached an AUC score
      of 0.65 (0.56, 0.70) and 0.73 (0.70, 0.75), respectively.  
   }
   \label{fig_roc}
\end{figure}

Receiver operating characteristics (ROC) results on the test set can be seen in Figure \ref{fig_roc}.
The test AUC for the video based network was given by 0.81 (0.77, 0.84),
for the 3D network trained from scratch and the pre-trained 2D network test AUCs
were given by 0.64 (0.56, 0.70) and 0.73 (0.70, 0.75), respectively. 

Sensitivity, specificity, positive and negative predicted values
(PPV and NPV respectively) and the $F_1$ score, given by the harmonic mean of precision and recall,
were computed for a threshold value of 0.50 in the output layer.
Test set results are shown in Table \ref{result_table}.
\begin{table}
   \centering
   \scalebox{0.8}{
      \begin{tabular}{l c c c}
         \hline
                     & \textbf{3D video pre-trained}  & \textbf{3D from scratch} & \textbf{2D ImageNet pre-trained}  \\
         \hline
         AUC         & 0.81 (0.02)  & 0.64 (0.05)  & 0.73 (0.02)  \\
         sensitivity & 0.75 (0.06)  & 0.67 (0.12)  & 0.84 (0.07)  \\
         specificity & 0.72 (0.09)  & 0.49 (0.09)  & 0.40 (0.13)  \\
         PPV         & 0.53 (0.07)  & 0.35 (0.03)  & 0.37 (0.04)  \\
         NPV         & 0.88 (0.02)  & 0.79 (0.05)  & 0.87 (0.03)  \\
         $F_1$ score & 0.62 (0.02)  & 0.45 (0.05)  & 0.51 (0.03)  \\
         \hline
      \end{tabular}
   }
   \caption{
      Test results for the three different models.
      Scores represent mean (std) values for the ten times each model was trained.
   }
   \label{result_table}
\end{table}

% -------------------------------- Discussion ---------------------------------------%

\section{Discussion}

From the three networks trained the video pre-trained C3D model performed best
with a test AUC score showing clear superiority over the two other models.

We associate the success of pre-training with two factors.
First of all, transfer learning is counteracting the problem of small data set sizes
by injection of knowledge prior to the actual learning task.
General benefit of natural imaging based pre-training for the medical domain
has been challenged by \citet{raghu2019transfusion},
accounting improvements only to application of over-parametrized models.
However, our training data set involved just a few hundred cases,
making even training of small networks difficult.
We therefore accredit the model improvement, in our case, to actual transferred knowledge
between the two domains.
Second, transfer learning improves generalization.
Different studies have shown that radiomic feature values are affected by CT scanners and scanning protocols
\cite{lu2016assessing, zhao2014exploring, larue2017influence, ger2018comprehensive}.
Convolutional neural networks are sensitive to such domain shifts \cite{hendrycks2019benchmarking, kurakin2016adversarial}.
\citet{hendrycks2019using} showed that transfer learning can improve model robustness,
which leads to better validation and testing results.

Additionally, we recognize improved performance of the 3D approach over the 2D approach.
Our results suggest that the third dimension contains essential information for HPV classification.

When tested on an external data set radiomic models reached AUC scores ranging between 0.70 and 0.82
\cite{bogowicz2020privacy, huang2019development}.
A crucial difference between radiomics and deep learning lies within the input information.
Radiomic methods apply filters on a predefined region of interest, typically the GTV, for feature generation.
HPV-positive OPCs, however, are known to be associated with regions
outside the tumor volume, e.g. cystic lymph nodes \cite{goldenberg2008cystic, begum2003detection}.
Therefore, radiomic approaches usually ignore important parts of the image.
Deep learning on the other hand does not require the delineation of the tumor volume,
leading to an advantage in HPV prediction.

Our work has proven the capability of deep learning models to predict patients HPV
status based on CT images from different institutions.
For clinical application further studies are required. 
This involves the implementation of more diverse training data to investigate
the impact on model generalizability.
Furthermore, the effect of data quality has to be examined.
Head and neck CT images are generally prone to artifacts, \citet{leijenaar2015external} analyzed a subset of
the \citest{tciaOPC} collection and found that 49\% of all cases contained visible artifacts.
They showed that involvement of artifacts plays a major role for HPV status prediction
by training a radiomics model which achieved AUC scores ranging between 0.70 and 0.80
depending on the inclusion of cases with artifacts in the training and test cohorts
\cite{leijenaar2018development}.

The 8$^{th}$ edition of the American Joint Committee of Cancer (AJCC) staging manual
defined HPV-positive OPCs as a independent entity and recommended testing by immunohistochemical
staining of p16, a surrogate marker of HPV \cite{lydiatt2017head}.
However, p16 overexpression is not exclusively linked to a ongoing carcinogenesis caused by a HPV infection,
leading to a low specificity of p16 testing with up to 20\% of OPC p16-positive cases being HPV-negative
\cite{wasylyk2013identification}.
Tests with higher specificity exist but are technically challenging.
The current gold standard test of E6/E7 mRNA detection by PCR remains labor-intensive and, hence, relatively expensive.
\cite{westra2014detection, qureishi2017current}. 
A imaging based tests on the other hand is easily applicable, non-invasive and also features time efficiency and low cost.
In order to improve total sensitivity and specificity it can also be used in combination
with other testing methods.

% --------------------------------  Conclusion --------------------------------------%

\section{Conclusion}

It was demonstrated that convolutional neural networks are able to classify the HPV status
of oropharyngeal cancer patients based on CT images from different cohorts.
The sports video clip based transfer learning approach performed best in comparison to two other CNN models.
Video based pre-training has the potential to improve deep learning on 3D medical data,
but further studies are needed for verification.

% --------------------------------- References --------------------------------------%

%\bibliographystyle{ieeetr}
\newpage
\printbibliography

%\newpage
%\input{./sections/supplementary}

\end{document}